%% file: bare_jrnl.tex
\documentclass[journal]{IEEEtran}

\usepackage{times}
\usepackage{soul}
\usepackage{url}
\usepackage[hidelinks]{hyperref}
\usepackage[utf8]{inputenc}
\usepackage[small]{caption}
\usepackage{graphicx}
\usepackage{amsmath}
\usepackage{amssymb}
\usepackage{booktabs}
\usepackage{algorithm}
\usepackage{algorithmic}
\usepackage{multirow}
\usepackage{epsfig}
\ifCLASSINFOpdf
\else
\fi
\begin{document}
%
\title{Cascaded Revision Network for Novel Object Captioning}
%
%
%

\author{Qianyu~Feng,
        Yu~Wu,
        Hehe~Fan,
        Chenggang~Yan,
        and~Yi~Yang
\thanks{Qianyu~Feng, Yu~Wu, Hehe~Fan and Yi~Yang are with Centre for Artificial Intelligence, University of Technology Sydney, NSW 2007, Australia. (e-mail:
qianyu.feng@student.uts.edu.au; Yu.Wu-3@student.uts.edu.au; Hehe.Fan@student.uts.edu.au; 
Yi.Yang@uts.edu.au) }
\thanks{Chenggang~Yan is with the Department of Automation, Hangzhou Dianzi University, Hangzhou, 310018, China (e-mail:
cgyan@hdu.edu.cn)}
}

\maketitle

\begin{abstract}
Image captioning, a challenging task where the machine automatically describes an image by sentences, has drawn significant attention in recent years.
Despite the remarkable improvements of recent approaches, however, these methods are built upon a large set of training image-sentence pairs.
The expensive labor efforts hence limit the captioning model to describe the wider world.
In this paper, we present a novel network structure, Cascaded Revision Network, which aims at relieving the problem by equipping the model with out-of-domain knowledge.
CRN first tries its best to describe an image using the existing vocabulary from in-domain knowledge.
Due to the lack of out-of-domain knowledge, the caption may be inaccurate or include ambiguous words for the image with unknown (novel) objects.
We propose to re-edit the primary captioning sentence by a series of cascaded operations.
We introduce a perplexity predictor to find out which words are most likely to be inaccurate given the input image.
Thereafter, we utilize external knowledge from a pre-trained object detection model and select more accurate words from detection results by the visual matching module.
In the last step, we design a semantic matching module to ensure that the novel object is fit in the right position.
By this novel cascaded captioning-revising mechanism, CRN can accurately describe images with unseen objects.
We validate the proposed method with state-of-the-art performance on the held-out MSCOCO dataset as well as scale to ImageNet, demonstrating the effectiveness of this method.
\end{abstract}


%
\IEEEpeerreviewmaketitle

\section{Introduction}

\IEEEPARstart{I}{mage} captioning has become a promising direction in the research for computer vision and language~\cite{DBLP:conf/cvpr/VinyalsTBE15,DBLP:conf/cvpr/JohnsonKF16,DBLP:conf/ijcai/chen18,DBLP:conf/ijcai/mao18,DBLP:conf/cvpr/AndersonHBTJGZ18,6544585,7778165,8003302}.
This task aims to automatically generate a natural and concrete description of an image.
Recent approaches based on the encoder-decoder structure have achieved encouraging performances on the image captioning task.
However, existing models could only describe the objects shown in the training image-caption pairs, which hinders the generalization of these models in real-world scenarios.
How to describe images with unseen objects is still a challenge for image captioning~\cite{DBLP:conf/cvpr/Hendricks2016CVPR,DBLP:conf/cvpr/Lu2018Neural,DBLP:conf/mm/WuZJY18}.

In this paper, we aim to alleviate this problem by equipping the image captioning model with out-of-domain knowledge.
Naturally, when seeing an unknown object, human search their memory and find the most similar object to describe it. 
For example, when seeing a ``zebra'', humans tend to project the features and the environment of the ``zebra'' and deduce:``It is something like a horse.''
If an additional knowledge database is available, \textit{e.g.}, picture flashcards or an internet search engine, human could look up similar objects and select the correct ``word'' to better describe the unknown object.
With out-of-domain knowledge, it is possible to learn the similarity and difference between a ``horse'' and a ``zebra'' and describe the unseen ``zebra'' with its correct name.


\begin{figure}[t]
\begin{center}
  \includegraphics[width=0.99\linewidth]{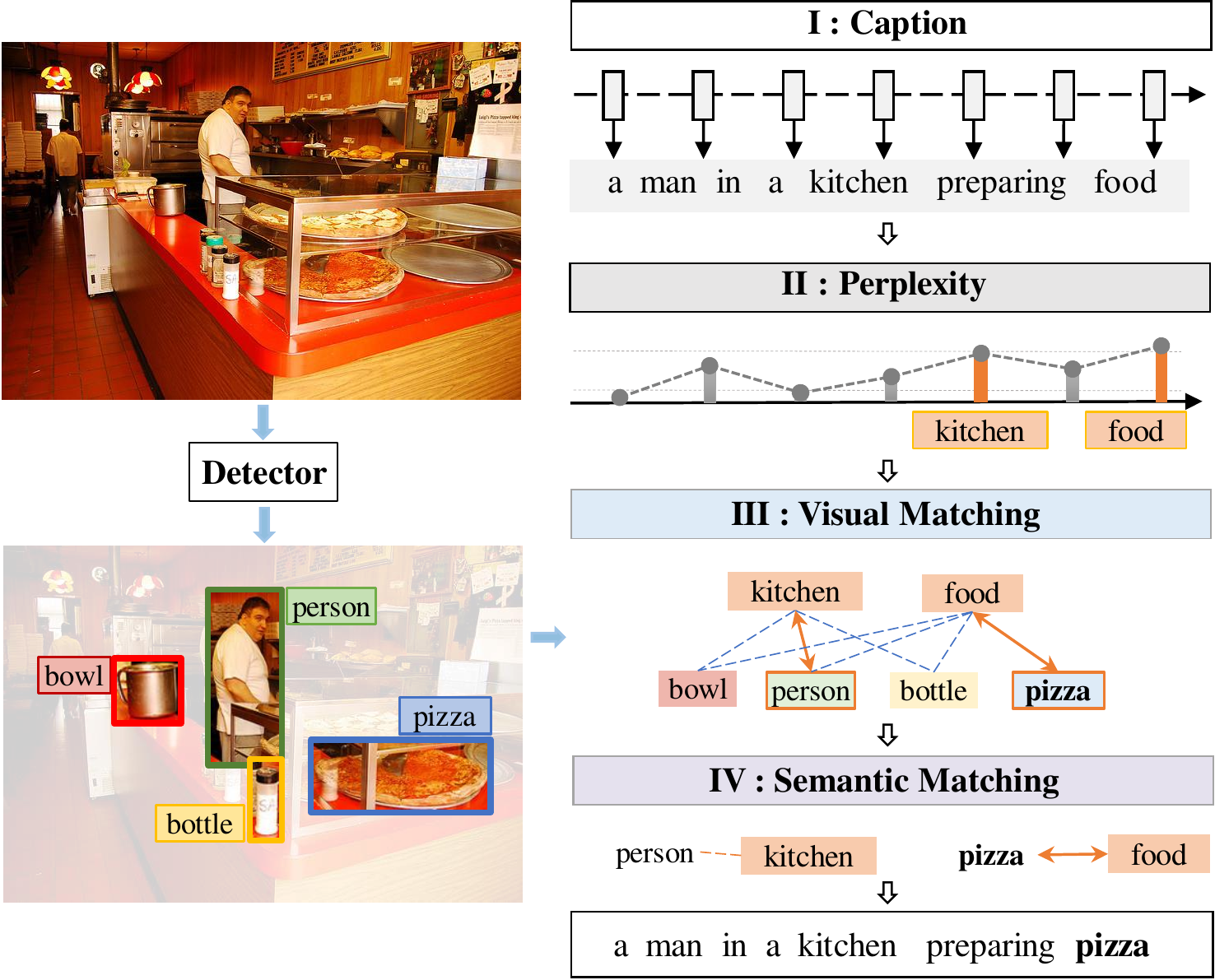}
  \caption{An example of novel object captioning by Cascaded Revision Network (CRN). 
  To leverage external knowledge, we use an object detector to provide out-of-domain information in the form of object-word pairs.
  CRN cascades a primary image captioner, a perplexity predictor, a visual matching module and a semantic matching module.
  {\bf I:} The image captioner never see the object ``pizza'' before and thus try its best to generate a caption with the known word ``food''.
  {\bf II:} The perplexity predictor is aware that ``kitchen'' and ``food'' are with high ambiguity.
  Note the ``kitchen'' is also predicted to be ambiguous, since it is not reliable for the model to trust ``kitchen'' given the input image.
  {\bf III:} The visual matching module generate replacement proposals by assigning a detected object to each of the ambiguous words: ``kitchen'' $\rightarrow$ ``person'' and ``food'' $\rightarrow$ ``pizza''.
  {\bf IV:} The word similarity matching module is designed to ensure the correctness of matching. It rejects the unreliable matching proposal (\textit{e.g.}, ``kitchen'' $\rightarrow$ ``person'') by measuring their word similarities.
  Only those similar word pairs (\textit{e.g.}, ``food'' $\rightarrow$ ``pizza'') are allowed to be replaced.
  By this captioning-revising mechanism, CRN is able to generate the accurate caption for those unseen object.
}
  \label{fig:pizza}
\end{center}

\end{figure}

In this paper, we introduce a novel framework called Cascaded Revision Network (CRN) for novel object captioning.
When describing an image with novel objects, the image captioner is first asked to try its best to characterize the image using existing in-domain knowledge.
To this end, the agent could choose synonyms or similar words in its vocabulary to describe unknown objects.
These synonyms or similar words can be ambiguous and even inaccurate due to the lack of out-of-domain knowledge.
We define a sentence generated by the image captioner with only in-domain vocabulary as a primary caption.

Imitating human-style describing an image with unseen objects, we design three cascaded operations to better revise the primary caption: 1) estimating the uncertainty of each word; 2) searching the external knowledge database for a better description; 3) embedding the out-of-domain object into the caption without breaking the grammar. 
In our CRN framework, the above sub-tasks are executed by the perplexity predictor, the visual matching module and the semantic matching module, respectively.
The perplexity predictor is designed to figure out the ambiguous words in the primary caption.
Specifically, the perplexity predictor checks each word in the primary caption and predicts a perplexity score for each word.
If a perplexity score is greater than the perplexity threshold, the corresponding word is considered as a candidate to be revised.
When predicting a word with high perplexity, the agent probably meets an object it is not familiar with or not sure about.
Thus, the agent needs to ask for help from external knowledge to generate more accurate words.
Besides, there are also cases when the agent is capable to caption the image based on its own knowledge. 
The agent can handle these cases with high confidence which means it does not need additional help.

Next, we leverage the external knowledge to find more accurate noun words for the cases with high perplexity.
In CRN, a pre-trained object detector is used to obtain all objects with their names in the image.
We then design the visual matching module to match the inaccurate words with detected objects.
We adopt the key-value memory mechanism to construct the communication between the captioning agent and the object detector.
Specifically, the agent uses the substitutes to query the memory according to the visual information of objects.
Then, the word corresponding to the object with the highest probability is selected to replace the substitute.
To this end, the visual matching module exploits an external object detector~\cite{DBLP:journals/pami/RenHG017} as out-of-domain knowledge.
In this way, the corresponding name of the selected object becomes a candidate to revise the primary caption.

However, the object detector is not always reliable to detect all the objects accurately. 
In this case, the visual matching module would generate a wrong matching proposal.
The semantic matching module is responsible for eliminating such incorrect visual match proposals.
Specifically, the semantic matching module measures the similarity between the ambiguous word and the object name with an out-of-domain word embedding.
The incorrect visual match could result in a small similarity and therefore be ignored.
By this cascaded captioning-revising mechanism, novel objects are described accurately in the final caption sentence.
An example of novel object captioning by CRN is illustrated in Figure~\ref{fig:pizza}.

Our proposed method turns out with results competitive with the current state-of-the-art performance on the held-out MSCOCO on the novel object captioning task. We also scale CRN to a larger dataset: ImageNet \cite{imagenet_cvpr09}. With more analysis, we reveal that our approach not only improves captioning with novel objects as well as images without novel objects. Finally, the main contributions of this paper are summarized as follows:
\begin{itemize}
    \item We propose a novel cascaded framework for novel object captioning by imitating how we human describe an image with an unseen object.
    At first, the model tries its best to generate a primary caption based on in-domain knowledge. We propose to gradually revise the primary captioning sentence by a series of cascaded operations.
    \item In the cascaded network, we develop a perplexity predictor, a visual matching module and a semantic matching module to revise the primary captioning.
    \item To our knowledge, we are the first to match the out-of-domain knowledge both visually and semantically to better combine it with the in-domain captions.
\end{itemize}

\section{Related Work}

{\bf Deep Image Captioning.}
Given an image, the goal of image captioning is to generate a natural and accurate sentence to describe the image.
Early approaches~\cite{DBLP:journals/pieee/YaoYLLZ10,DBLP:conf/nips/OrdonezKB11} composed image captions via slot filling which separate the object recognition and the language template generation. 
These approaches may generate natural sentences but less related to the visual contents.
Deep Learning has elevated the performance of captioning models with images and videos. Most of related work~\cite{mao2014deep,DBLP:conf/cvpr/KarpathyL15,DBLP:conf/cvpr/DonahueHGRVDS15,DBLP:conf/icml/XuBKCCSZB15,DBLP:conf/iccv/MaoWYWHY15,DBLP:conf/aaai/li17,DBLP:conf/aaai/mun17,DBLP:conf/aaai/xin19,7509595,8221809} follow a multimodal framework which combines CNN~\cite{DBLP:conf/iclr/SimonyanZ14a}, ~\cite{DBLP:conf/cvpr/HeZRS16}
and RNN like Long Short-Term Memory (LSTM)~\cite{DBLP:journals/neco/HochreiterS97} and Gated Recurrent Unit (GRU)~\cite{DBLP:conf/emnlp/ChoMGBBSB14}.
Visual features in high-level with semantic information are first extracted by the CNN encoder, while the RNN decoder predicts the description word by word according to visual features.
However, these methods do not consider the situation where a large number of unseen objects exist in the images.

{\bf Zero-shot Learning.}
With the booming development of techniques in computer vision, lack of well-labeled data becomes the bottleneck of performance. 
Image-paired sentences is in large scarcity and the label tagging of captioning costs much more than other tasks. 
Zero-shot learning is a good solution to resolve the out-of-domain adaptation for models with limited knowledge. There has been a recent surge on the zero-shot tasks \cite{DBLP:journals/pami/LampertNH14,DBLP:conf/cvpr/XianLSA17,DBLP:conf/cvpr/DingSF17} which aims to recognize objects unseen during the training stage. Many two-stage approaches \cite{DBLP:conf/nips/FromeCSBDRM13} are proposed to first capture the attributes of the unseen objects, then infer the class label with the most similar set of features.

{\bf Novel Object Captioning.}
The novel object captioning task attracts increasing attention recently. The problem exists in how to leverage the unpaired image and semantic data \cite{DBLP:conf/cvpr/LinMBHPRDZ14} to better describe the unseen objects.
A few works are carried out to address this captioning task. The Deep Compositional Captioner (DCC) is proposed by \cite{DBLP:conf/cvpr/Hendricks2016CVPR}, a pilot work to put forward the task of novel objects captioning. DCC~\cite{DBLP:conf/cvpr/Hendricks2016CVPR} combine visual groundings of lexical units to generate descriptions about objects which are not present in caption corpora (paired image-sentence data), but are present in object recognition datasets (unpaired image data) and text corpora (unpaired text data). 
Novel Object Captioner (NOC) \cite{DBLP:conf/cvpr/VenugopalanHRMD16} is introduced as an end-to-end framework training the object classification, language model and the captioning jointly. The detection model is integrated with the language sequence model by copying detection results into the prediction out of the RNN-based decoder model to alleviate the gap between novel objects with the captioning model in \cite{DBLP:conf/cvpr/YaoCVPR17}.
An approach is proposed in \cite{DBLP:conf/cvpr/Lu2018Neural} to generate language template along with slots and the corresponding region in the image at first. Then objects are fit into the slots by recognizing the region with a detection model. But they have to manually define the category of the novel object with an existing one when captioning. What is more, the categories are still not well defined and limit the concepts to similar visual out-looking which is too idealized. For example, ``man'' and ``woman'' are classified as ``person'', while ``car'', ``bus'' and ``truck'' belong to three different classes. 
A placeholder is used in \cite{DBLP:conf/mm/WuZJY18} to take place of the novel objects which generalize the concept of novel object but also lost information of the current object. 
These methods rely too much on visual detection. The results are limited to the detection model and less likely to select small objects. They neglected the original lexical context information. To the best of our knowledge, our model is the first captioning model with self-awareness and two-way revision mechanism.

{\bf Summary.}
In a nutshell, the proposed method focuses on generating accurate caption of images with novel objects. With the cascaded revision mechanism, CRN exploits the out-of-domain knowledge provided by the object detector and better embeds the novel object with the in-domain captions. With the setting of \textit{pseudo objects}, CRN is able to distinguish unknown objects from correct ones. What's more, the cascaded visual matching and semantic matching ensures the combination of out-of-domain objects with the in-domain descriptions.

\section{The Proposed Approach}
\begin{figure*} 
\begin{center}
\includegraphics[width=\linewidth]{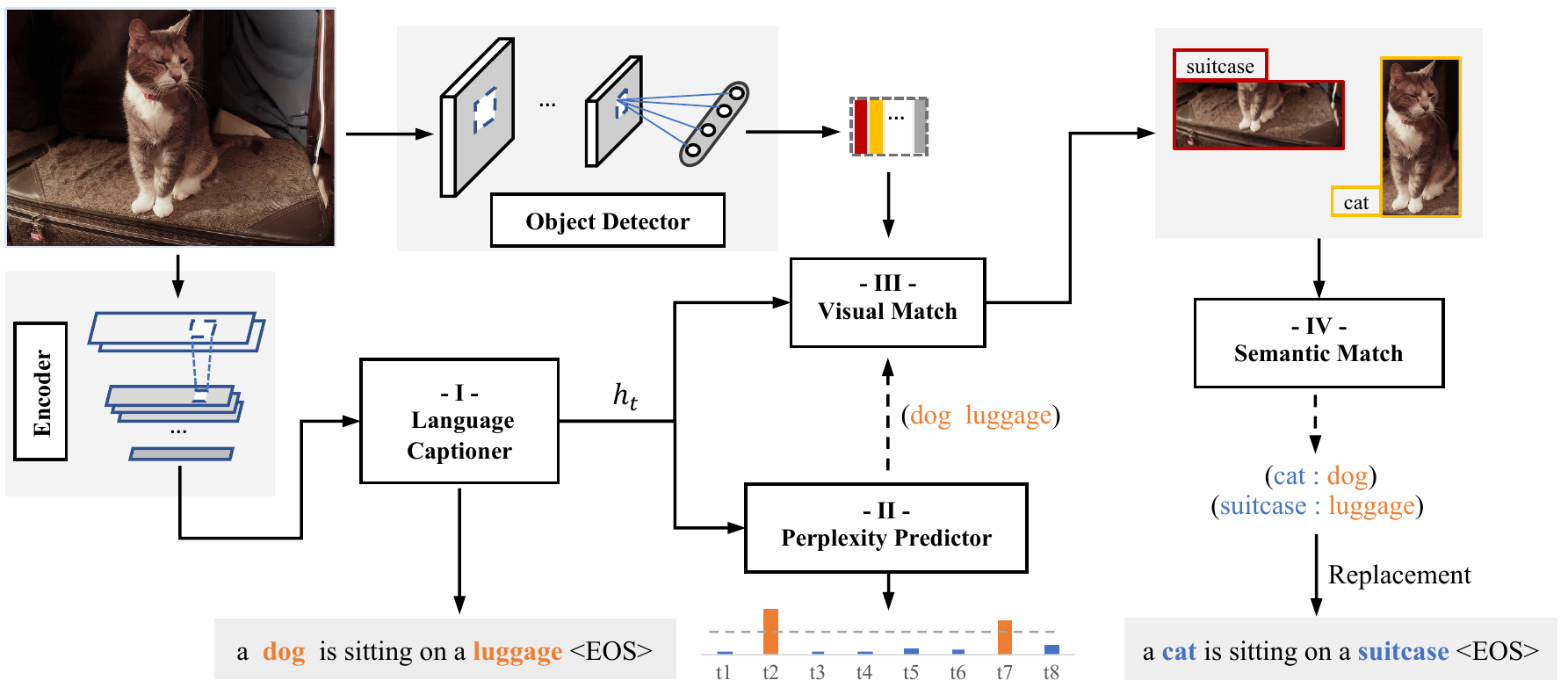}
\end{center}
  \vspace{.1in}
  \caption{The overview of the evaluation stage of the proposed framework. ``cat'' and ``suitcase'' are novel objects to the captioning model that never exist in training. At stage I, the language captioner tries its best to generate a sentence based on the existing vocabulary. We propose to revise this sentence to get a more accurate description by a series of cascaded operations.
  At stage II, the perplexity predictor is applied to find ambiguous words, \textit{e.g.}, ``dog'' and ``luggage''.
  At stage III and IV, we match and replace these words with detected objects based on both visual and semantic similarity.
  Finally, detected objects ``cat'' and ``suitcase'' are fit in the right positions in the sentence.}
\label{fig:main}
\end{figure*}

The Cascaded Revision Network(CRN) is designed to better embed the out-of-domain objects into the in-domain captions.
In this section, we first introduce the traditional image captioning model in Section~\ref{subsec:Image-Captioning-Model}.
We then show how CRN describes images with novel objects in Section~\ref{subesec:CRN}. The full framework of CRN is illustrated in Figure~\ref{fig:main}.

\subsection{Image Captioning Model}
\label{subsec:Image-Captioning-Model}
The main task of an image captioning model is to generate a natural language sentence to describe the image, while maintaining the semantic grammar of the sentence. 
Given an image $I$ and the ground truth caption $w =  \{w_1, w_2,...,w_T\}$, the objective of the captioning model is to minimize
\begin{equation}\label{eq1}
\begin{split}
L	& = -\mathrm{log}\ p(y|I) = -\mathrm{log}\ p(w_1, w_2, ..., w_T|I) \\
 	& = -\mathrm{log}\ \prod_{t=1}^T p(w_t|w_1, w_2, ..., w_{t-1}, I) \\
    & = -\sum_{t=1}^T \mathrm{log}\ p(w_t|w_1, w_2, ..., w_{t-1}, I).
\end{split}
\end{equation}
Eq.~\ref{eq1} aims to maximize the likelihood of each word in the ground-truth caption. 
Usually, the term $p(w_t|w_1, ..., w_{t-1}, I)$ is modeled by a Long Short-Term Memory (LSTM)~\cite{DBLP:journals/neco/HochreiterS97} that takes $I$ as its initial state $h_0$:
\begin{equation}\label{eq2}
p(\cdot|w_1, w_2, ..., w_{t-1}, h_0), h_t = LSTM(w_{t-1}, h_{t-1}),
\end{equation}
where $w_0$ is the start symbol \textless START\textgreater. 
What's more, the distribution $p(\cdot|w_1, w_2, ..., w_{t-1}, h_0)$ is a parametric function of $h_t$. 
LSTM first generates the current hidden state $h_t$ and then emits the distribution by a fully-connected layer according to $h_t$. 
For simplicity, we use $\pi(\cdot|h_t)$ to denote this distribution: 
\begin{equation}\label{eq3}
\pi(\cdot|h_t) = p(\cdot|w_1, w_2, ..., w_{t-1}, h_0).
\end{equation}
The current word is generated by
\begin{equation}\label{eq4}
w_t = \arg \max_w\ \pi(\cdot|h_t).
\end{equation}
During training, the previous ground-truth words are given.
When conducting the evaluation, the previous ground-truth words are unavailable and are generated by maximum likelihood estimation (MLE). 

\subsection{Cascaded Revision Network}
\label{subesec:CRN} 
CRN aims to alleviate the problem of novel object captioning by equipping the model with out-of-domain knowledge. 
To exploit out-of-domain knowledge, CRN adopts a captioning-revising mechanism. 
Following~\cite{DBLP:conf/cvpr/Lu2018Neural,DBLP:conf/mm/WuZJY18}, in this paper, we use the out-of-domain knowledge provided by an object detector and a word embedding look-up table.
CRN contains four cascaded modules: a primary image captioner, a perplexity predictor, a visual matching module and a semantic matching module.
With the setting of \textit{pseudo objects}, CRN learns to distinguish the ambiguous words inconsistent with the images.

\subsubsection{Image Captioner}
The main challenge of this task is that the model has no prior knowledge of novel objects. 
In this case, the captioning model will predict a word based on the visual looking or the semantic context. 
Specifically, the captioner describes an image with its existing vocabulary to generate a primary caption. 
Ambiguous or even inaccurate words may be used when describing unknown objects. 
Several words are assigned as novel objects during the training of captioner based on the encoder-decoder framework described in \ref{subsec:Image-Captioning-Model}.
Here, we denote words in the vocabulary of the image-paired captions as $V_c$. The objects neither in the images nor the captions are novel objects denoted as $O_u$. To simulate the existence of novel objects, objects are selected from the vocabulary $V_c$ to be replaced in the captions which are denoted as $O_i$. Objects $\in O_i$ act as the role of novel objects during training which the model has never seen. 
We replace objects $\in O_i$ with \textit{pseudo objects}.
With the open-source pretrained embeddings, each object $\in O_i$ is paired with its most similar word $\in V_c$ which acts as \textit{pseudo object}. The \textit{pseudo object} and its corresponding object $\in O_i$ form a pair of inaccurate description of an object. The word similarity is measured with the cosine metric between the word embeddings. Furthermore, in order to inform the captioner about the existence of \textit{pseudo object}, we design an additional novel label $\hat {n}$ of each word $w$ to indicate whether it is a novel object or not: 
\begin{equation}
    \hat{n} =\begin{cases} 1, w \in O_i\\ 
    0, otherwise.\end{cases}
\label{novel-label}
\end{equation}

Another embedding function $\phi_n$ is adopted to embed the novel label into the input of captioner. At time step $t$, the input vector of the captioner $x_t$ is the concatenation of the embedding of $w_{t-1}$ and its novel label $\hat{n}_{t-1}$:
\begin{equation}
\begin{aligned}
    x_t & =\left[\phi_e(w_{t-1}), \phi_n(\hat{n}_{t-1})\right] \\
               & =\left[W_e I^w_{t-1}, W_n I^n_{t-1}\right],
\end{aligned}
\end{equation}
where $W_e \in \mathbb{R}^{N_v \times D_e}$ is the word embedding matrice of the vocabulary $V_c$. $N_v$ is the number of the vocabulary. $D_e$ denotes the dimension of embedding. $W_n \in \mathbb{R}^{2 \times D_e}$ denotes learnable weight matrice of the novel label $\hat{n}_t$. $I^w_{t-1}$ and $I^n_{t-1}$ are the corresponding one-hot encoding of $w_{t-1}$ and $\hat{n}_{t-1}$. 
With the input vector $x_t$, the output hidden state of captioner is given by: 
\begin{equation}
    h_t = w^{T}_{h} \tanh{(W_s x_t+W_z h_{t-1})},
\end{equation}
where $w^{T}_{h}, W_s, W_z$ are weights to be learned. At each time step, the distribution of the conditional probabilities over all possible words $\in V_c$ is: 
\begin{equation}
    p_t = softmax (W_p h_t+b_p),
\end{equation}
where $W_p, b_p$ are learned weights and biases.

\subsubsection{Perplexity Predictor}
To revise the primary caption, the perplexity predictor is designed to figure out the ambiguous words in it. The intuition behind the proposed method is to enable the captioner to justify whether the word is consistent with the image or not. Thus, it is aware of the ambiguity of its outputs. We here define the level of ambiguity as semantic perplexity. In information theory, perplexity is a measurement of how well a model predicts a sample. The perplexity of the current output of captioner is calculated using the hidden state of captioner. The function of the perplexity predictor is designed as:
\begin{equation}
    m_t = \sigma (W_m h_t+b_m),
\end{equation}
where $W_m, b_m$ are learned weights and biases for this layer. $\sigma$ is the sigmoid activation of confidence probability. A threshold $\tau_p$ is adopted here. If $m_t$ surpasses $\tau_p$, it indicates that the current prediction is not with enough confidence. All outputs with high perplexity will become regarded as inaccurate words and will probably be replaced with by a matched object in the next revision steps.

With the image captioner and the perplexity predictor introduced above, the corresponding objective cross entropy loss function is:
\begin{equation}
\begin{aligned}
    &L_{cap}(w_{1:t-1}, I;\theta) = \\
    &-\dfrac {1}{T} (\sum ^{T}_{t=1} \log p(w_t| w_{1:t-1}) + \sum ^{T}_{t=1} \log p(m_t|w_{1:t-1})).
\end{aligned}
\end{equation}

\subsubsection{Visual Matching Module}
The visual matching module is responsible for acquiring objects in the image with the knowledge of the detector and generate replacement proposals based on visual similarity.
To introduce novel objects out-of domain into the image captioner, we employ an freely available pretrained object detection model $M_d$. 
Thus, we can take advantage of $M_d$ to detect objects in the image which are furher used to revise the inaccurate words in the primary caption. 
The extracted visual features $V_d \in \mathbb{R}^{N_o\times D_v}$. $N_o$ is the number of detected objects. $D_v$ is the dimension of visual feature. The predicted class labels $O_d\in\mathbb{R}^{1\times N_d}$ of the objects can also be obtained from $M_d$. $N_d$ is the number of target classes of $M_d$. We extract the visual features of objects from the ROI pooling layer of $M_d$ following \cite{DBLP:conf/cvpr/JohnsonKF16}. The objects are chosen according to the prediction scores given by the object detection model.
With the hidden state $h_t$ of captioner at time step $t$, the visual similarity between the current feature and features of all detected objects can be calculated as:
\begin{equation}
    S_t = V_d h_t.
\end{equation}
Then we address the probabilities over all classes of $M_d$ at time $t$:
\begin{equation}
    O_t = S_t O_d,
\end{equation}

Each inaccurate word will be matched with a detected object which is regarded as a candidate to be put in the final caption. For the matching between the output of captioner and the feature of detected objects, the objective for training this module is defined as:
\begin{equation}
    L_{det}(h_t;\theta) = - \dfrac {1}{N_d} \sum ^{N_d}_{i=1} \hat{n}_t \log p(o_t|h_t),
\end{equation}
where $N_{d}$ is the number of detected objects at time step $t$, $n_t$ is used as the mask of the current ground truth word which is defined in Eq \eqref{novel-label}.
These three modules of CRN are jointly trained during the training of CRN.

\subsubsection{Semantic Matching Module} 
Simply replacing the inaccurate words with the visually matched objects may break the semantic structure of the sentences. Besides, due to the limitation of the compressed features, objects with salient features tend to be matched with a high frequency. It is observed that many ambiguous words are matched with the same detected object while some are not relative semantically. Therefore, we elevate the quality of revision by employing the semantic matching as the last step. 

With the selected objects from the detection model, the word similarity is calculated with the pretrained word embedding look-up. It is noticed that there are some words which are composed of two words cannot be found in the Glove embedding, \textit{e.g.}, ``hot dog'', ``hair drier'', etc. In this case, to prevent manual intervention, we simply average the embeddings of the two words. The cosine similarity is used to measure the distance between the novel objects and the caption words. The word with the largest word similarity is replaced by the detected object.

Finally, the full framework of CRN is proposed to deal with the captioning of images with novel objects. With the different modules cascaded in the model, each module is optimized with a sub-goal. 
The gap between the novel object and the existing knowledge is represented by the perplexity of the prediction which simulates the process of thinking before the description.

\section{Experiments}
We start by describing the setups of this task and our experiments. Then, the results of our methods and the state-of-the-art methods in history are compared on the held-out MSCOCO dataset. Furthermore, several ablation studies are carried out with competitive results to prove the effectiveness and reliability of our proposed method.

\input{tables/1-allmethods.tex}
\subsection{Experimental Settings}
MSCOCO is a widely used benchmark for many tasks including image captioning \cite{DBLP:conf/cvpr/LinMBHPRDZ14}. 
The held-out subset of the MSCOCO dataset following \cite{DBLP:conf/cvpr/Hendricks2016CVPR,DBLP:conf/cvpr/VenugopalanHRMD16,DBLP:conf/cvpr/YaoCVPR17} are used as the training set in our experiments. In \cite{DBLP:conf/cvpr/Hendricks2016CVPR}, eight classes of MSCOCO objects are chosen.
None of the 8 classes is included in the captioning in the training split set, but all of them are in the evaluation split set. We follow the same setting of training, validation and test split in \cite{DBLP:conf/cvpr/Hendricks2016CVPR} in order to generate comparable captioning results.

\begin{figure}[t]
\begin{center}
  \includegraphics[width=0.9\linewidth]{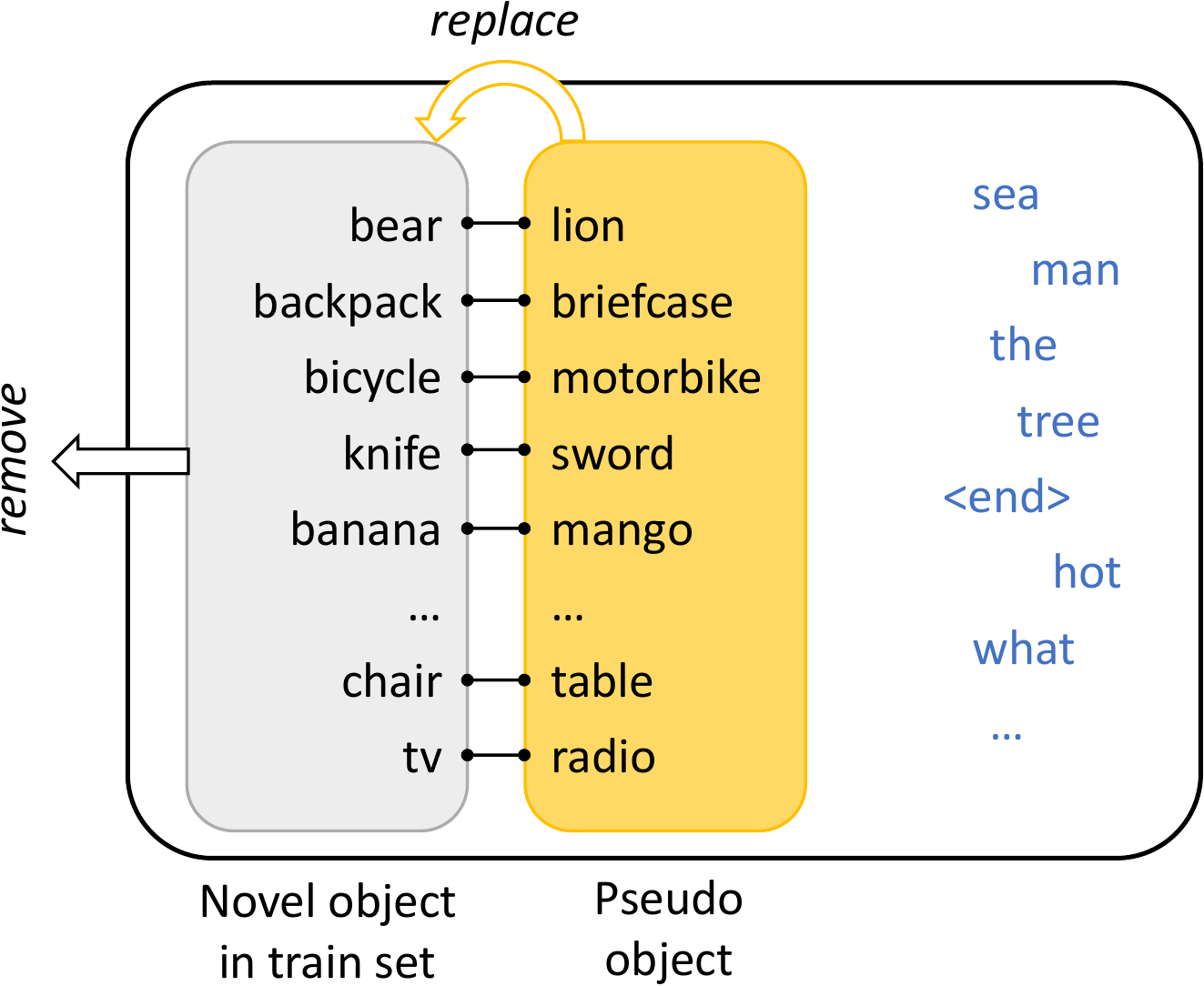}
  \vspace{.1in}
  \caption{Example of words in the in-domain vocabulary. The selected novel objects in the training captioning sentences are replaced with the pseudo objects. Each pseudo object is matched by comparing the semantic similarity with the selected novel object, \textit{e.g.}, ``banana'' is replaced by ``mango'' in our training. We use the pseudo objects to train the matching and replacement mechanism in our framework.
}
  \label{fig:pseudo}
\end{center}
\end{figure}

{\bf Pseudo object processing.}
All classes except the eight held-out classes in MSCOCO are chosen as novel objects $O_i$ in the train set which are replaced with \textit{pseudo objects} in the in-domain vocabulary. To select \textit{pseudo object} of each novel object $\in O_i$, we employ the open source pretrained embedding weights of Glove following \cite{DBLP:conf/cvpr/Hendricks2016CVPR,DBLP:conf/cvpr/YaoCVPR17} with the dimension of 300. For example, ``umbrella"$\rightarrow$``parasol", ``zebra"$\rightarrow$``horse", ``sandwich"$\rightarrow$``burger", etc. We stress that we have not used any other semantic data or description for these objects, neither do we manually change any word. The detail of the replacement is shown in Figure~\ref{fig:pseudo}. It comes out the plural format of the word tends to be the most similar word to itself, \textit{e.g.}, ``sandwiches" to ``sandwich". It is meaningless if we use word ``sandwiches" to take place of ``sandwich", as they refer to the same object. 

{\bf Experiment details.}
We apply a 16-layer VGG pretrained on ImageNet following \cite{DBLP:conf/cvpr/Hendricks2016CVPR,DBLP:conf/cvpr/VenugopalanHRMD16,DBLP:conf/cvpr/YaoCVPR17} as the image encoder in our model. Parameters of the encoder are frozen during the training. The features output by layer fc7 are used as the representation of the image and fed into the language decoder. The dimension of the image feature is 4,096. 
In order to introduce the novel objects into the final captions, a popular open-source pre-trained Faster-RCNN model \cite{DBLP:journals/pami/RenHG017} is adopted to detect and crop the objects in an image following \cite{DBLP:conf/EMNLP/AndersonFJG16a,DBLP:conf/cvpr/Lu2018Neural,DBLP:conf/mm/WuZJY18}. Then, we reuse the VGG Net mentioned above to extract visual features of the detected objects. 
The pre-trained detection model is released by \cite{DBLP:conf/cvpr/HuangRSZKFFWSG016}\footnote{\url{https://github.com/tensorflow/models}}, which is trained on all the 80 classes of objects in the MSCOCO detection dataset. 
We adopt the LSTM as the decoder with one layer and its dimension is 1024.

{\bf Compared approaches.}
To evaluate on the held-out MSCOCO, results of our proposed method are compared with DCC \cite{DBLP:conf/cvpr/Hendricks2016CVPR}, NOC \cite{DBLP:conf/cvpr/VenugopalanHRMD16}, LSTM-C \cite{DBLP:conf/cvpr/YaoCVPR17}, Base+T4 \cite{DBLP:conf/EMNLP/AndersonFJG16a}, NTB+G \cite{DBLP:conf/cvpr/Lu2018Neural} and DNOC \cite{DBLP:conf/mm/WuZJY18} to demonstrate the competitiveness. During the methods, NTB+G and DNOC do not use the additional semantic data. We follow the same zero-shot setting in our experiments. 
Furthermore, the results of several ablation versions of the proposed model are compared and discussed. 
In order to prove the advantage of CRN not only exists in the novel object captioning, we also evaluate F1 scores of other known objects $\in W_{paired}$ in Table \ref{tab:objects-sub-1}. 

\input{tables/3-nocs-I.tex}


\subsection{Compared to the state-of-the-art methods}
Captions are being evaluated with the widely-used COCO caption evaluation tool. 
For the task of novel object captioning, only the METEOR metric is not enough for the evaluation. Sentences with good grammar can obtain high scores even without mentioning the novel objects. 
The caption is deemed accurate only if the correct novel object appears at least once in the sentence. 
The results of our proposed model with the F1 scores to measure the performance on novel objects and METEOR are presented in Table \ref{tab:allmethods} along with all state-of-the-art methods on the held-out MSCOCO dataset. 
The F1-scores of all novel objects surpass the best state-of-the-art result while the average F1 score achieves 64.08\% (6.16\% higher than 57.92\%). 
It is observed that methods \cite{DBLP:conf/cvpr/VenugopalanHRMD16, DBLP:conf/cvpr/YaoCVPR17} with external text data including the novel objects perform better on several objects than the proposed CRN. However, there may always be objects novel to the captioner that it never learned from the text nor the image.
Our METEOR score is lower than LSTM-C with GloVe \cite{DBLP:conf/cvpr/YaoCVPR17} a little bit. Nevertheless, our experiments are carried out based on the zero-shot setting. what is more, all of the eighty classes of objects are novel to our model. It is explainable that captioning model can better describe the context of the known objects than objects never seen before. 

\input{tables/2-ablation.tex}

\input{tables/imagenet.tex}

\begin{figure*}
\begin{center}
\includegraphics[width=\linewidth]{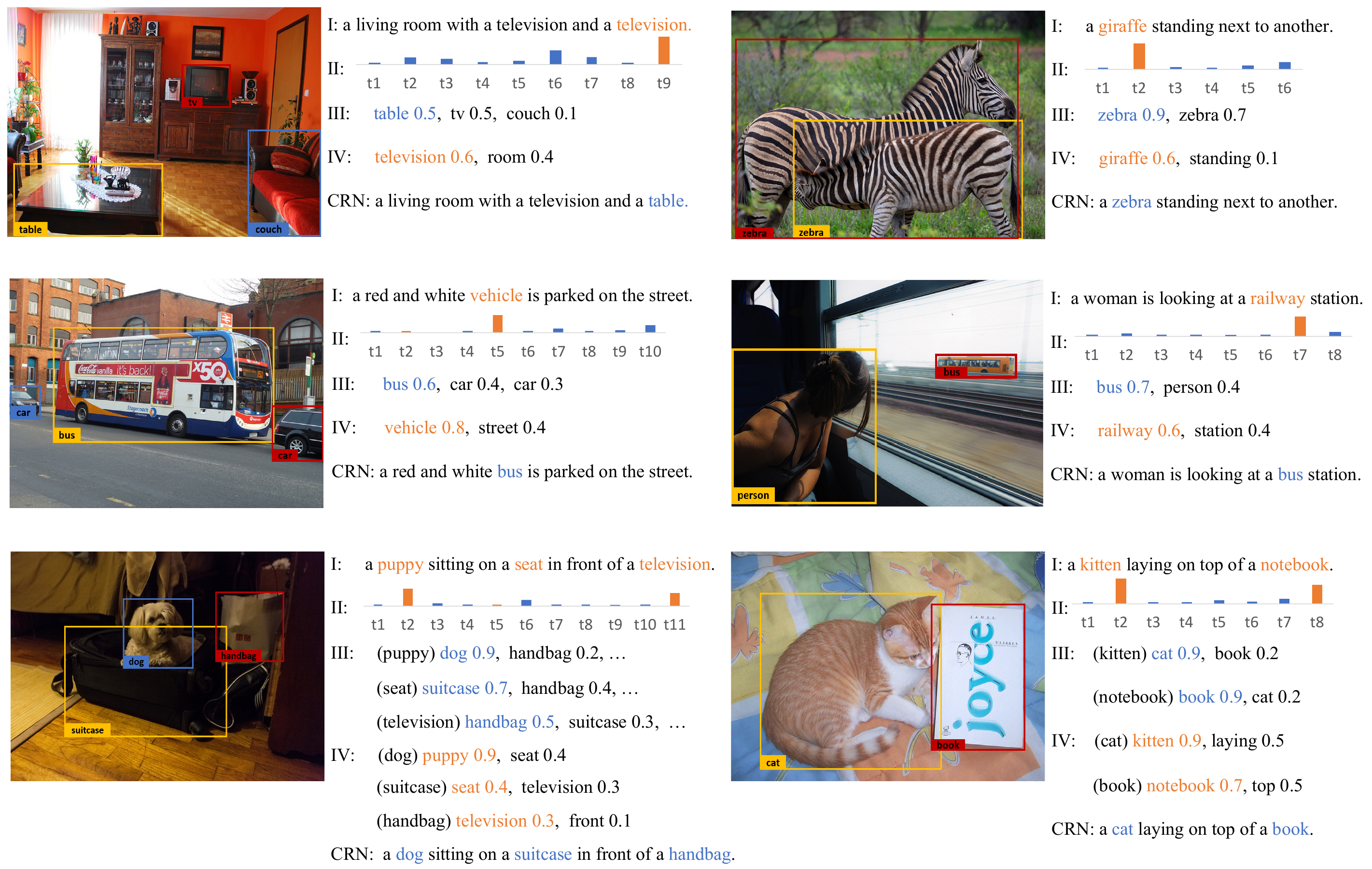}
\end{center}
  \caption{Captions generated by CRN on the held-out MSCOCO where images contains unseen objects. Boxes with colors are object candidates proposed by a pre-trained object detector. Sentences with tag ``I'' are the initial captions generated by CRN-I only with the in-domain vocabulary. At step ``II'', the perplexity predictor outputs the perplexity of each word. Then each ambiguous word with perplexity higher than the threshold will be matched with a detected object at step ``III''. At the last step, the matched object(s) will be fit into the primary caption by selecting the word with the highest semantic similarity.}
  \vspace{.1in}
\label{fig:pics}
\end{figure*}

\subsection{Ablation Studies}
The ablation studies are conducted on the held-out MSCOCO dataset with the same setting mentioned above.
We compare different ablation versions of CRN to prove the effectiveness of the sub-modules: the perplexity prediction and the revision of objects. Results are listed in Table \ref{tab:first-ablation}. 

{\bf CRN I} is CRN only with the captioner which knows nothing about the novel objects as LRCN. Thus, the F1 score is 0. The existence of the \textit{pseudo novel object} leads to the drop of METEOR score. 
{\bf CRN I+II} (perplexity predictor) adds the second task: predicting the perplexity of each word. If the perplexity goes beyond the threshold $\tau_p$, the word will be replaced by a detected object randomly selected from the results of the detection model. It brings a significant rise in the F1 score. The METEOR also increases from 18.12 to 19.65. The threshold $\tau_p$ is set as 0.15 in our experiments learned by the model. 
{\bf CRN w/o II} is CRN without the perplexity predictor. As the average number of words above the perplexity threshold in the training stage is 1.7 per sentence, we choose two positions in the sentence to replace the detected object matched with the two-way matching of visual similarity and word similarity. 
It shows that \textit {CRN I+II} is better on METEOR than \textit {CRN w/o II} which indicates the value of the perplexity predictor. 
The average F1 score of \textit {CRN w/o II} is 53.31\%, 8.01\% higher than \textit {CRN I+II}.
{\bf CRN w/o IV} (semantic matching) is CRN without the matching of word similarity. Objects are matched only with the features from the language decoder and visual features of objects detected. The F1 average score increases from 45.30\% to 56.32\%. 
{\bf CRN w/o III} (visual matching) objects are matched only with word similarity which outperforms \textit {CRN w/o IV} by 5.76\% on F1 score. With full stages, our model is able to capture features of the unknown objects on visual out-looking and semantic context which composes more accurate captions about the image.
Furthermore, in order to show the advantage of the proposed model not only exist in the novel object captioning, we also evaluate F1 scores of other words $\in W_s$. Our model is also able to generate accurate descriptions of known objects. F1 scores on a different group of known objects are listed in Table \ref{tab:objects-sub-1}. It turns out that the performance on these objects is also quite qualitative. Figure~\ref{fig:pics} shows some examples of image captioning results with novel objects. 

\begin{figure}
\begin{center}
\includegraphics[width=\linewidth]{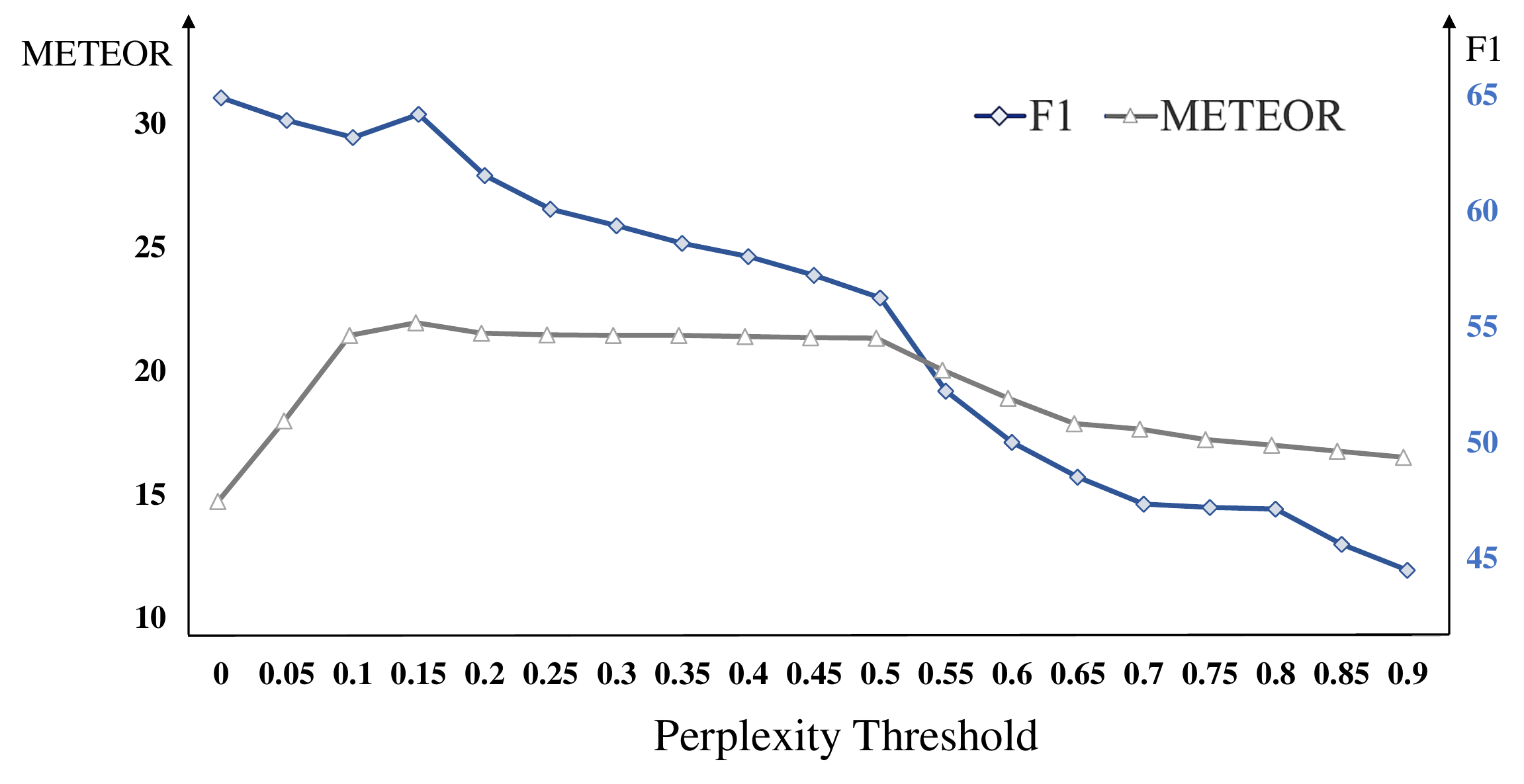}
\end{center}
  \caption{The effect of perplexity threshold on the performance of CRN on the held-out MSCOCO. The left y-axis is the scale of METEOR and the right y-axis measures the average F1 score.}
  \vspace{.1in}
\label{fig:threshold}
\end{figure}

{\bf Threshold of Perplexity.} We present the performance of F1 score and METEOR along with the change of threshold of perplexity in Figure~\ref{fig:threshold}. When the threshold is 0, the F1 score achieves quite high but with a low meteor. It indicates that the objects detected by the detection model in the image are replaced into the caption while it destroys the grammar and structure of the sentence. When the threshold is between (0, 0.15), METEOR both get higher, while F1 score drops slightly. It indicates that the threshold limits the ambiguous words area while the number of objects replaced into the sentence decreased. After that, the F1 score goes down when the threshold is larger than 0.15. The METEOR score also decreases and drops more when threshold goes beyond 0.5.

{\bf Scale to Larger Dataset.} The proposed CRN takes advantage of an expertised detector to introduce novel objects. Considering the out-of-MSCOCO objects, a detector with larger vocabulary should be adopted. Hence, we report the performance of using the detector pretrained on Visual Genome \cite{krishnavisualgenome} and scaling CRN to ImageNet. Results of additional experiments are reported in Table \ref{tab:secode-ablation}.

\section {Conclusion}

In this paper, we present a novel cascaded framework CRN to deal with captioning with novel objects. To overcome the gap between existing knowledge and objects out-of-domain, the captioner in CRN is able to be aware of what is ambiguous or unknown to itself. 
Furthermore, with a two-way matching mechanism, the unknown object can be better matched and fit in the caption.
At a higher level, our proposed method decouples the captioning of novel objects to two sub-tasks: what is the novel object and where to put the novel object. By applying the two-way matching, CRN better integrates the out-of-domain knowledge both visually and semantically. 

\bibliographystyle{IEEEtran}
\bibliography{references}

\begin{IEEEbiographynophoto}{Qianyu Feng}
received the M.S. degree in Shanghai Jiao Tong
University, China, in 2018. She is currently a Ph.D.
student in University of Technology Sydney, Australia. Her research interests include image captioning
and domain adaptation.
\end{IEEEbiographynophoto}

\begin{IEEEbiographynophoto}{Yu Wu}
 received the B.S. degree in Shanghai Jiao Tong
University, China, in 2015. He is currently a Ph.D.
student in University of Technology Sydney, Australia. His research interests include language navigation
and person re-identification.
\end{IEEEbiographynophoto}


\begin{IEEEbiographynophoto}{Hehe Fan}
received the M.S. degree in Huazhong University of Science and Technology, China, in 2015. He is currently a Ph.D.
student in University of Technology Sydney, Australia. His research interests include video classification
and person re-identification.
\end{IEEEbiographynophoto}

\begin{IEEEbiographynophoto}{Chenggang Yan}
received the B.S. degree in Computer Science from Shandong University in 2008 and
Ph.D. degree in Computer Science from the Institute of Computing Technology, Chinese Academy
of Sciences in 2013. Now he is the Director of
Intelligent Information Processing Lab in Hanzhou
Dianzi Univeristy. Before that, he was an assistant
research fellow in Tsinghua University. His research
interests include intelligent information processing,
machine learning, image processing, computational
biology and computational photography.
\end{IEEEbiographynophoto}

\begin{IEEEbiographynophoto}{Yi Yang}
received the Ph.D. degree in computer science from Zhejiang University, China, in 2010. He is
currently a professor with University of Technology
Sydney, Australia. He was a Post-Doctoral Research
with the School of Computer Science, Carnegie Mellon University, USA. His current research interest includes machine learning and its applications to multimedia content analysis and computer vision, such
as multimedia indexing and retrieval, surveillance
video analysis and video semantics understanding.
\end{IEEEbiographynophoto}




\end{document}

%% file: tables/1-allmethods.tex
\begin{table*}[!t]
\small
\setlength{\tabcolsep}{11pt}
\centering
\begin{tabular}{@{}l|c|cccccccc|c@{}}
\hline
Method & METEOR & F$_\text{bottle}$ & F$_\text{bus}$ & F$_\text{couch}$ & F$_\text{microwave}$ 
& F$_\text{pizza}$ & F$_\text{racket}$ & F$_\text{suitcase}$ & F$_\text{zebra}$ & F$_\text{average}$ \\ 
\hline\hline


         DCC~\cite{DBLP:conf/cvpr/Hendricks2016CVPR}    & 21      & 4.63   & 29.79  & 45.87  & 28.09  & 64.59  & 52.24  & 13.16  & 79.88  & 39.78  \\


         NOC*~\cite{DBLP:conf/cvpr/VenugopalanHRMD16}   & 21.32   & 17.78  & 68.79  & 25.55  & 24.72  & 69.33  & 55.31  & 39.86  & 48.79 & 48.79  \\

         LSTM-C~\cite{DBLP:conf/cvpr/YaoCVPR17} & 22    & 29.07  & 64.38  & 26.01  & 26.04  & 75.57  & 66.54  & 55.54  & \textbf{92.03}  & 54.40\\

         \begin{tabular}[c]{@{}l@{}}LSTM-C*\end{tabular} 
        
                                            & 23 & 29.68  & 74.42  & 38.77  & 27.81  & 68.17  & \textbf{70.27} & 44.76 & 91.40 & 55.66 \\


         
         Base+T4$^\dagger$  \cite{DBLP:conf/EMNLP/AndersonFJG16a}  & \textbf{23.6} & 16.3 & 67.8 & 48.2 & 29.7 & 77.2 & 57.1 & 49.9 & 85.7 & 54.0\\

         NBT+G \cite{DBLP:conf/cvpr/Lu2018Neural}  & 22.8        & 7.1    & 73.7   & 34.4 & \textbf{61.9} & 59.9 &  20.2 & 42.3  & 88.5  & 48.5 \\ 
        
         DNOC \cite{DBLP:conf/mm/WuZJY18}   & 21.57          & 33.04   & 76.87 & 53.97 & 46.57 & 75.82 & 32.98 & \textbf{59.48} & 84.58 & 57.92\\
         \textbf{CRN (ours)}    & 21.31 & \textbf{38.05} & \textbf{78.40} & \textbf{55.93} & 53.76 & \textbf{81.43} & 62.02 & 57.69 & 85.38  & \textbf{64.08}\\
\hline

\end{tabular}
\vspace{.05in}
\caption{Comparison with the state-of-the-art methods on F1 score and METEOR score. All results are generated with image feature extracted by VGG-16 \protect\cite{DBLP:conf/iclr/SimonyanZ14a} and without beam search. $^\dagger$ is method with Resnet-based CNN and beam search. F1-score values are reported in format of percentage (\%). * indicates training with pretrained Glove word embedding weights.}
\vspace{.1in}
\label{tab:allmethods}
\end{table*}

%% file: tables/3-nocs-I.tex
\begin{table}[!tb]
\footnotesize
\small
\setlength{\tabcolsep}{3.3pt}
\centering

\begin{tabular}{@{}l ccccccc @{}}
\toprule
Method & F$_\text{bear}$      & F$_\text{cat}$      & F$_\text{dog}$      & F$_\text{elephant}$      & F$_\text{horse}$      & F$_\text{motorcycle}$      & F$_\text{average}$      \\ \midrule
LRCN~\cite{DBLP:conf/cvpr/DonahueHGRVSD14}   & 66.23           & 75.73          & 53.62          & 65.49               & 55.20            & 71.45           & 64.62       \\
DNOC~\cite{DBLP:conf/mm/WuZJY18}   & 62.86           & 87.28          & 71.57          & 77.46               & 71.20            & 77.59           & 74.66       \\
CRN    & 60.38           & 86.74          & 74.04          & 81.41               & 75.36            & 78.39           & {\bf 76.05}     
\\ \bottomrule
\\
\end{tabular}
\caption{Comparison on F1 scores of pseudo novel objects from subset 1 with baseline LRCN and DNOC.}
\vspace{.1in}
\label{tab:objects-sub-1}
\end{table}

%% file: tables/2-ablation.tex
\begin{table}[!tb]
\footnotesize
\small
\setlength{\tabcolsep}{25pt}
\centering

\begin{tabular}{@{}lcc@{}}
\toprule
Method               & METEOR & F$_\text{average}$ \\ \midrule
LRCN~\cite{DBLP:conf/cvpr/DonahueHGRVSD14}                 & 19.33  & 0                  \\
CRN I                & 18.24  & 0                  \\
CRN I + II           & 19.65  & 45.30              \\
CRN w/o II           & 19.26  & 53.31              \\
CRN w/o IV           & 20.85  & 56.32              \\
CRN w/o III          & 21.01  & 62.08              \\ \bottomrule
\\
\end{tabular}
\caption{Ablation studies on each component of CRN.}
\vspace{.1in}
\label{tab:first-ablation}
\end{table}


%% file: tables/imagenet.tex
\begin{table}[!tb]
\footnotesize
\small
\setlength{\tabcolsep}{22pt}
\centering
\begin{tabular}{@{}lccc@{}}
\toprule
Method       & Novel   & F$_\text{average}$     & Acc           \\ \midrule
NOC~\cite{DBLP:conf/cvpr/VenugopalanHRMD16}          & 69.08   & 15.63  & 10.04         \\
LSTM-C~\cite{DBLP:conf/cvpr/YaoCVPR17}       & 72.08   & 16.39  & 11.83         \\
CRN          & \textbf{77.92}   & \textbf{19.5}   & \textbf{16.34}         \\ \bottomrule
\\
\end{tabular}
\caption{Comparison with state-of-the-art methods on ImageNet.}
\label{tab:secode-ablation}
\end{table}

%% file: bare_jrnl.bbl
\begin{thebibliography}{10}
\providecommand{\url}[1]{#1}
\csname url@samestyle\endcsname
\providecommand{\newblock}{\relax}
\providecommand{\bibinfo}[2]{#2}
\providecommand{\BIBentrySTDinterwordspacing}{\spaceskip=0pt\relax}
\providecommand{\BIBentryALTinterwordstretchfactor}{4}
\providecommand{\BIBentryALTinterwordspacing}{\spaceskip=\fontdimen2\font plus
\BIBentryALTinterwordstretchfactor\fontdimen3\font minus
  \fontdimen4\font\relax}
\providecommand{\BIBforeignlanguage}[2]{{%
\expandafter\ifx\csname l@#1\endcsname\relax
\typeout{** WARNING: IEEEtran.bst: No hyphenation pattern has been}%
\typeout{** loaded for the language `#1'. Using the pattern for}%
\typeout{** the default language instead.}%
\else
\language=\csname l@#1\endcsname
\fi
#2}}
\providecommand{\BIBdecl}{\relax}
\BIBdecl

\bibitem{DBLP:conf/cvpr/VinyalsTBE15}
O.~Vinyals, A.~Toshev, S.~Bengio, and D.~Erhan, ``Show and tell: {A} neural
  image caption generator,'' in \emph{CVPR}, 2015.

\bibitem{DBLP:conf/cvpr/JohnsonKF16}
J.~Johnson, A.~Karpathy, and L.~Fei{-}Fei, ``Densecap: Fully convolutional
  localization networks for dense captioning,'' in \emph{CVPR}, 2016.

\bibitem{DBLP:conf/ijcai/chen18}
H.~Chen, G.~Ding, Z.~Lin, S.~Zhao, and J.~Han, ``Show, observe and tell:
  Attribute-driven attention model for image captioning,'' in \emph{IJCAI},
  2018.

\bibitem{DBLP:conf/ijcai/mao18}
Y.~Mao, C.~Zhou, X.~Wang, and R.~Li, ``Show and tell more: Topic-oriented
  multi-sentence image captioning,'' in \emph{IJCAI}, 2018.

\bibitem{DBLP:conf/cvpr/AndersonHBTJGZ18}
P.~Anderson, X.~He, C.~Buehler, D.~Teney, M.~Johnson, S.~Gould, and L.~Zhang,
  ``Bottom-up and top-down attention for image captioning and {VQA},'' in
  \emph{CVPR}, 2018.

\bibitem{6544585}
P.~V.~K. {Borges}, N.~{Conci}, and A.~{Cavallaro}, ``Video-based human behavior
  understanding: A survey,'' \emph{IEEE Transactions on Circuits and Systems
  for Video Technology}, 2013.

\bibitem{7778165}
R.~V. H.~M. {Colque}, C.~{Caetano}, M.~T.~L. {de Andrade}, and W.~R.
  {Schwartz}, ``Histograms of optical flow orientation and magnitude and
  entropy to detect anomalous events in videos,'' \emph{IEEE Transactions on
  Circuits and Systems for Video Technology}, 2017.

\bibitem{8003302}
K.~{Kang}, H.~{Li}, J.~{Yan}, X.~{Zeng}, B.~{Yang}, T.~{Xiao}, C.~{Zhang},
  Z.~{Wang}, R.~{Wang}, X.~{Wang}, and W.~{Ouyang}, ``T-cnn: Tubelets with
  convolutional neural networks for object detection from videos,'' \emph{IEEE
  Transactions on Circuits and Systems for Video Technology}, 2018.

\bibitem{DBLP:conf/cvpr/Hendricks2016CVPR}
L.~Anne~Hendricks, S.~Venugopalan, M.~Rohrbach, R.~Mooney, K.~Saenko, and
  T.~Darrell, ``Deep compositional captioning: Describing novel object
  categories without paired training data,'' in \emph{CVPR}, 2016.

\bibitem{DBLP:conf/cvpr/Lu2018Neural}
J.~Lu, J.~Yang, D.~Batra, and D.~Parikh, ``Neural baby talk,'' in \emph{CVPR},
  2018.

\bibitem{DBLP:conf/mm/WuZJY18}
Y.~Wu, L.~Zhu, L.~Jiang, and Y.~Yang, ``Decoupled novel object captioner,'' in
  \emph{ACM MM}, 2018.

\bibitem{DBLP:journals/pami/RenHG017}
S.~Ren, K.~He, R.~B. Girshick, and J.~Sun, ``Faster {R-CNN:} towards real-time
  object detection with region proposal networks,'' \emph{{IEEE} Trans. Pattern
  Anal. Mach. Intell.}, 2017.

\bibitem{imagenet_cvpr09}
J.~Deng, W.~Dong, R.~Socher, L.-J. Li, K.~Li, and L.~Fei-Fei, ``{ImageNet: A
  Large-Scale Hierarchical Image Database},'' in \emph{CVPR}, 2009.

\bibitem{DBLP:journals/pieee/YaoYLLZ10}
B.~Z. Yao, X.~Yang, L.~Lin, M.~W. Lee, and S.~C. Zhu, ``{I2T:} image parsing to
  text description,'' \emph{IEEE}, 2010.

\bibitem{DBLP:conf/nips/OrdonezKB11}
V.~Ordonez, G.~Kulkarni, and T.~L. Berg, ``Im2text: Describing images using 1
  million captioned photographs,'' in \emph{NIPS}, 2011.

\bibitem{mao2014deep}
J.~Mao, W.~Xu, Y.~Yang, J.~Wang, Z.~Huang, and A.~Yuille, ``Deep captioning
  with multimodal recurrent neural networks (m-rnn),'' in \emph{ICLR}, 2015.

\bibitem{DBLP:conf/cvpr/KarpathyL15}
A.~Karpathy and F.~Li, ``Deep visual-semantic alignments for generating image
  descriptions,'' in \emph{CVPR}, 2015.

\bibitem{DBLP:conf/cvpr/DonahueHGRVDS15}
J.~Donahue, L.~A. Hendricks, S.~Guadarrama \emph{et~al.}, ``Long-term recurrent
  convolutional networks for visual recognition and description,'' in
  \emph{CVPR}, 2015.

\bibitem{DBLP:conf/icml/XuBKCCSZB15}
K.~Xu, J.~Ba, R.~Kiros, K.~Cho, A.~C. Courville, R.~Salakhutdinov, R.~S. Zemel,
  and Y.~Bengio, ``Show, attend and tell: Neural image caption generation with
  visual attention,'' in \emph{ICML}, 2015.

\bibitem{DBLP:conf/iccv/MaoWYWHY15}
J.~Mao, X.~Wei, Y.~Yang, J.~Wang, Z.~Huang, and A.~L. Yuille, ``Learning like a
  child: Fast novel visual concept learning from sentence descriptions of
  images,'' in \emph{ICCV}, 2015.

\bibitem{DBLP:conf/aaai/li17}
L.~Li, S.~Tang \emph{et~al.}, ``Image caption with global-local attention,'' in
  \emph{AAAI}, 2017.

\bibitem{DBLP:conf/aaai/mun17}
J.~Mun, M.~Cho, and B.~Han, ``Text-guided attention model for image
  captioning,'' in \emph{AAAI}, 2017.

\bibitem{DBLP:conf/aaai/xin19}
X.~Wang, J.~Wu, D.~Zhang, Y.~Su, and W.~Y. Wang, ``Learning to compose
  topic-aware mixture of experts for zero-shot video captioning,'' in
  \emph{AAAI}, 2019.

\bibitem{7509595}
S.~{Coşar}, G.~{Donatiello}, V.~{Bogorny}, C.~{Garate}, L.~O. {Alvares}, and
  F.~{Brémond}, ``Toward abnormal trajectory and event detection in video
  surveillance,'' \emph{IEEE Transactions on Circuits and Systems for Video
  Technology}, 2017.

\bibitem{8221809}
J.~{Wu} and H.~{Hu}, ``Cascade recurrent neural network for image caption
  generation,'' \emph{Electronics Letters}, 2017.

\bibitem{DBLP:conf/iclr/SimonyanZ14a}
K.~Simonyan and A.~Zisserman, ``Very deep convolutional networks for
  large-scale image recognition,'' in \emph{ICLR}, 2015.

\bibitem{DBLP:conf/cvpr/HeZRS16}
K.~He, X.~Zhang, S.~Ren, and J.~Sun, ``Deep residual learning for image
  recognition,'' in \emph{CVPR}, 2016.

\bibitem{DBLP:journals/neco/HochreiterS97}
S.~Hochreiter and J.~Schmidhuber, ``Long short-term memory,'' \emph{Neural
  Computation}, 1997.

\bibitem{DBLP:conf/emnlp/ChoMGBBSB14}
K.~Cho, B.~van Merrienboer \emph{et~al.}, ``Learning phrase representations
  using {RNN} encoder-decoder for statistical machine translation,'' in
  \emph{EMNLP}, 2014.

\bibitem{DBLP:journals/pami/LampertNH14}
C.~H. Lampert, H.~Nickisch, and S.~Harmeling, ``Attribute-based classification
  for zero-shot visual object categorization,'' \emph{{IEEE} Trans. Pattern
  Anal. Mach. Intell.}, 2014.

\bibitem{DBLP:conf/cvpr/XianLSA17}
Y.~Xian, C.~H. Lampert, B.~Schiele, and Z.~Akata, ``Zero-shot learning - {A}
  comprehensive evaluation of the good, the bad and the ugly,'' in \emph{CVPR},
  2017.

\bibitem{DBLP:conf/cvpr/DingSF17}
Z.~Ding, M.~Shao, and Y.~Fu, ``Low-rank embedded ensemble semantic dictionary
  for zero-shot learning,'' in \emph{CVPR}, 2017.

\bibitem{DBLP:conf/nips/FromeCSBDRM13}
A.~Frome, G.~S. Corrado, J.~Shlens \emph{et~al.}, ``Devise: {A} deep
  visual-semantic embedding model,'' in \emph{NIPS}, 2013.

\bibitem{DBLP:conf/cvpr/LinMBHPRDZ14}
T.~Lin, M.~Maire \emph{et~al.}, ``Microsoft {COCO:} common objects in
  context,'' in \emph{CVPR}, 2014.

\bibitem{DBLP:conf/cvpr/VenugopalanHRMD16}
S.~Venugopalan, L.~A. Hendricks, M.~Rohrbach, R.~J. Mooney, T.~Darrell, and
  K.~Saenko, ``Captioning images with diverse objects,'' in \emph{CVPR}, 2017.

\bibitem{DBLP:conf/cvpr/YaoCVPR17}
T.~Yao, Y.~Pan, Y.~Li, and T.~Mei, ``Incorporating copying mechanism in image
  captioning for learning novel objects,'' in \emph{CVPR}, 2017.

\bibitem{DBLP:conf/EMNLP/AndersonFJG16a}
P.~Anderson, B.~Fernando, M.~Johnson, and S.~Gould, ``Guided open vocabulary
  image captioning with constrained beam search,'' in \emph{EMNLP}, 2017.

\bibitem{DBLP:conf/cvpr/HuangRSZKFFWSG016}
J.~Huang, V.~Rathod, C.~Sun \emph{et~al.}, ``Speed/accuracy trade-offs for
  modern convolutional object detectors,'' in \emph{CVPR}, 2016.

\bibitem{DBLP:conf/cvpr/DonahueHGRVSD14}
J.~Donahue, L.~A. Hendricks, S.~Guadarrama \emph{et~al.}, ``Long-term recurrent
  convolutional networks for visual recognition and description,'' in
  \emph{CVPR}, 2015.

\bibitem{krishnavisualgenome}
R.~Krishna, Y.~Zhu, O.~Groth, J.~Johnson, K.~Hata, J.~Kravitz \emph{et~al.},
  ``Visual genome: Connecting language and vision using crowdsourced dense
  image annotations,'' \emph{International Journal of Computer Vision}, 2016.

\end{thebibliography}
